\documentclass[12pt]{article}
\usepackage{apacite}

\usepackage[utf8]{inputenc}
\usepackage[american]{babel}
\usepackage[margin=1in]{geometry}
\usepackage{setspace} 
\usepackage{times}
\usepackage{fancyhdr}
\usepackage{authblk}
\usepackage{tabularx}

\usepackage{longtable}
\usepackage{booktabs}
\usepackage{multirow}
\usepackage{enumitem}
\usepackage{graphicx}
\usepackage{multicol}
\usepackage{array}
\usepackage[justification=centering]{caption} 
\usepackage[table]{xcolor}
\definecolor{lightgray}{gray}{0.9}
\usepackage{url}
\NewCommandCopy{\oldtabularx}{\tabularx}
\renewcommand*{\tabularx}{\rowcolors{1}{}{lightgray}\oldtabularx}
\usepackage{tikz} \usetikzlibrary{positioning}
\usetikzlibrary{arrows.meta} \usetikzlibrary{fit}

\usepackage{tabularray}
\newcounter{tblrowidx}

\newcommand{\getrowcolor}{\stepcounter{tblrowidx}\ifodd\value{tblrowidx}\gfancybackground{white}\else\gfancybackground{gray!15}\fi}
\let\citeNP\shortciteNP
\let\cite\shortcite

\title{\textbf{TeamCAMS: An Open-Source Research Platform for Studying Human Behaviour in Human-AI Teams}}
\date{}

\author[1]{Amos Brocco}
\author[2]{Alain Chavaillaz}
\author[2]{Andreas Sonderegger}
\author[2]{Juergen Sauer}

\affil[1]{University of Applied Sciences and Arts of Southern Switzerland, Department of Innovative Technologies, East Campus USI-SUPSI, Via la Santa 1, 6962 Lugano-Viganello, Switzerland}
\affil[2]{University of Fribourg, Department of Psychology, Rue P.-A.-de-Faucigny 2, 1700 Fribourg, Switzerland}

\begin{document}

\maketitle
\thispagestyle{fancy}

\begin{abstract}
In this article, we present TeamCAMS (Cabin Air Management System), a collaborative work environment for simulating human-AI (artificial intelligence) interaction for scientific research. The article outlines how several psychological theories guided the development of this multiple-task simulation. Modelling a process control environment, previous versions of TeamCAMS have already been used in empirical studies to address a wide range of research questions (e.g., comparing different forms of automation, evaluating impact of automation reliability, effects of stress on multiple-task performance). Outlining the technical possibilities offered by TeamCAMS, the article points out how its latest version offers researchers the possibility of addressing a set of new research questions including problems associated with teamwork (e.g., within-team conflict, distributed teamwork) and human-AI interaction. Finally, we will outline how this simulation environment can be enhanced further still to address research questions in new fields (e.g., automation of leadership). To promote transparency, reproducibility, and further development, TeamCAMS is made available to the research community under an open-source license.
\end{abstract}


\section{Introduction and overview}

\subsection{Need for simulation of complex work domains}
For many decades, there has been a need to simulate complex work domains by means of lab-based research tools. A primary motive for developing lab-based research tools has been due to the general difficulties of measuring performance objectively in real work domains \cite{alluisi_methodology_1967,avolio_leadership_2009,salthouse_determinants_1996}, which has been of some concern in ergonomics and psychology for some time. Furthermore, there has been a need to conduct experimental research to establish cause-effect relationships between variables rather than merely relying on correlational studies (e.g., \citeNP{antonakis_making_2010,bos_fundamental_2001}). Such experimental research can be conducted more easily with a lab-based simulation of a real work environment than in a field setting, hence providing an important complement to field research. For some years, the advancement of technologies has made such scientific endeavours easier to accomplish since this technical advancement allows more complex work domains to be modelled \cite{elsmore_synwork1_1994}. For example, the use of complex algorithms allows the implementation of advanced forms of automation, which adapt to the performance and behaviour of human operators \cite{parasuraman_humans_2008}. Overall, these technical developments may provide improved opportunities for researchers to model complex work with a view to gaining a better understanding of human behaviour in work environments of such complexity.  

There has been a long tradition in psychology to use computer-based simulations to model complex work \cite{brehmer_experiments_1993}. This refers to complex technical work environments (e.g., \citeNP{wiczorek_supporting_2014}) but also to complex non-technical work environments such as managerial decision-making \cite{dorner_strategic_1993}. Examples of the latter are being in command of a fire brigade \cite{dorner_strategic_1993} or managing a town as a mayor \cite{curral_leadership_2016,dorner_lohhausen_1994}. The endeavours to model complex work domains may even extend to complex sociotechnical systems, which also encompasses the organisational structure combined with the technical system \cite{hettinger_modelling_2015}. This was also the idea behind managerial decision-making in the tradition of microworld research, modelling scientific or social phenomena to learn and consolidate domain-specific skills and test ideas \cite{papert_mindstorms_1980}, like policymaking \cite{morecroft_system_1988} or gene inhibition process \cite{dunbar_concept_1993}. 

A particularly challenging work domain represents (industrial) process control (e.g., chemical plants, power plants, heating systems). Process control is generally characterised by three features \cite{brehmer_experiments_1993}. First, there are underlying dynamic processes that evolve autonomously, resulting in a constantly changing problem space. The problem space keeps evolving even in the absence of operator intervention. Second, there are closely coupled variables, requiring the concurrent completion of several tasks by the human operator (e.g., responding to alarms, diagnosing system faults, stabilising disturbed system states). Third, there is a certain level of opaqueness in the system, with some system variables not being directly accessible. This requires drawing inferences from those system variables, which are directly observable. These three features make process control a very challenging work domain, which has many implications for human factors and ergonomics (e.g., system design, training issues, diagnostic aids, predictive aids).

\subsection{Simulations for modelling complex technical work domains}
Over the years, researchers have developed a considerable number of work-related simulations, covering a range of different application areas. Some of these were developed in the context of the military for training and research purposes. Examples include the Robotic Non-commissioned Officer (NCO) \cite{parasuraman_adaptive_2009}, in which a human operator supervises under different levels of automation multiple uninhabited air and ground vehicles under high task load, or the Distributed Dynamic Decision-making Synthetic team Task (DDD) which is about simulating military command and control for a group of operators who deploy assets and coordinate resources with team mates \cite{kleinman_normative-descriptive_1998}. Firefighting is another application area with several simulations being created. This includes Fire Chief \cite{omodei_fire_1995} and DESSY \cite{brehmer_dynamic_1991}, which have been developed to investigate complex decision-making behaviour of fire brigade commanders. As a further example, C3Fire (Command, Control, and Communication Forest-Fire; \citeNP{johansson_c3fire_2003}) allows training the complexity and dynamics of collaborative work of fire fighters (team decision making and team situation awareness).  

One of the most prominent application areas is the process control environment. Rancor \cite{ulrich_rancor_2017,ulrich_hunter_2022} and COSSplay (Computerized Operator Support System; \citeNP{ulrich_cossplay_2016}) focus on process control in the complex work environments of a nuclear power plant while M-TOPS (Multi-Task Operator Performance Simulation; \cite{wiczorek_supporting_2014} simulates those tasks for chemical plants and CHESS (Central HEating System Simulation; \citeNP{wastell_time_2009}) for central heating systems. In the transportation domain, ATREIDES (Adaptive Train Research Enhanced Information Display \& Environment Simulator; \citeNP{naweed_designing_2013}) models the work environment of a train driver in a high-speed passenger train. Rail Signalling \cite{krehl_cognitive_2014} was developed to assess the cognitive resources required to carry out tasks in the context of railway signalling while watch keeping on ship bridges was simulated with CABOT \cite{sauer_designing_2003}. Further examples from different work domains include harvesting tasks \cite{sanchez_understanding_2014}, or even the complex tasks of managerial decision-making in the form of a mayor governing a small town \cite{dorner_lohhausen_1994}. 

This overview shows that simulations have been developed for a wide range of application areas with different objectives. The simulations differ greatly in terms of complexity and type of tasks, but also in terms of the stated objectives for which they were developed. The training of certain professional groups is often referred to as the primary goal of the development of simulations (e.g. firefighters using C3Fire or military and emergency services using DDD to train strategic thinking and team coordination), but the simulations were often also developed to study very specific constructs (e.g. to evaluate human reliability and situation awareness in high-risk environments with Rancor or to evaluate decision making process with Fire Chief and DESSY). All simulations have in common that they record the behaviour of the operators and thus allow an objective measurement of performance. Furthermore, they attempt to recreate real-world conditions with a satisfactory degree of realism.

The safe environment provided by these simulations (whether in the form of PC-based simulations or high-fidelity simulators) cannot only be used for research purposes but provide good training opportunities for industry (e.g. training airline pilots or process control operators). Typically, high-fidelity simulations are used for training in complex environments, whereas low-fidelity simulations are more often use in research \cite{dorner_lohhausen_1994,kluge_being_2012}. There are several advantages of using such simulation environments (e.g., designing tailor-made scenarios that can be adapted to specific training requirements, operating in a safe environment while the real work environment is very dangerous; \citeNP{nerdinger_arbeits-_2019}).  

\section{Theoretical underpinning of the simulation environment}
One goal of the present article is to demonstrate how the development of a simulation environment can be guided by psychological theory. TeamCAMS (Team Cabin Air Management System) will serve as a good example for that purpose. Please note that the current version has been termed TeamCAMS (due to its capacity to model teamwork) but previous versions have used different names (e.g., AutoCAMS; see below for more details). The following three theoretical models have notably influenced its development: Model of compensatory control mechanisms \cite{hockey_compensatory_1997}, automation taxonomy of \cite{sheridan_human_1987}, and the conceptual distinction between static, adaptive and adaptable automation (e.g., \citeNP{parasuraman_mitigating_2008}). 

The model of compensatory control mechanisms \cite{hockey_compensatory_1997} has guided the development of CAMS. Hockey’s model describes how individuals maintain performance under conditions of stress, fatigue, or high workload by reallocating cognitive resources. The model suggests that when demands exceed available resources, people engage in compensatory strategies to preserve performance on tasks they perceive as most important. This may result in performance decrements, most often on tasks considered being of lesser importance (i.e., secondary tasks), while performance on primary tasks is generally maintained because of possibly dire consequences of performance decrements. Operators may also make strategic adjustment to their system management behaviour (e.g., by taking calculated risks in the form of reduced information sampling). Finally, operators may activate additional cognitive-energetical resources to cope with the increased work demands. The design of CAMS has considered these adaptation processes in that it allows assessing these different human responses to sub-optimal working conditions. The feature of a multiple-task environment allows assessing different cognitive tasks, which are defined as primary and secondary tasks. Furthermore, operator behaviour such as information sampling behaviour and system control actions is recorded. Finally, subjective operator responses (e.g. mental effort) can be assessed by means of embedded questionnaires.  

The automation taxonomy by \cite{sheridan_human_1987} has guided the design of CAMS. The taxonomy proposes ten levels of automation, which define the relationship between the human and the machine with regard to their role in sharing responsibilities during task completion. The ten levels range from full manual control (level 1; i.e. human carries out all actions, receiving no support from the machine) through intermediate levels of automation (e.g., level 6; i.e. machine decides but the decision can be vetoed by the human) to a fully automated system (level 10; machine acts autonomously, ignoring the human). The design of CAMS has currently implemented the first six levels of automation as proposed by the taxonomy, with the implementation of the other four levels remaining an option for further development of the simulation environment. The closed modelling of CAMS on the automation taxonomy allows a good testing bed for examining the consequences of implementing certain automation levels on human-machine-system performance and operator state. The classic automation taxonomy by Sheridan and Verplank has triggered the development of further automation models, of which certain elements were also incorporated in CAMS design. A good example of this represents the theoretical distinction between the phases ‘information analysis’ and ‘action implementation’ in automation design \cite{parasuraman_model_2000}, which was then became the subject of empirical research by \cite{manzey_human_2012}.  

An important theoretical advancement in the automation literature represents the distinction between static, adaptive and adaptable automation (e.g., \citeNP{inagaki_adaptive_2008,miller_designing_2007}). The last two represent some flexible form of automation, which allows changing between different levels. The decision to change the level of automation can be initiated by a human (adaptable automation) or by the machine (adaptive automation). The consequences of choosing one form over the other have been intensely debated (e.g., see \citeNP{calhoun_adaptable_2022}) because of its impact on performance and operator state. The questions surrounding its impact become even more complex because there are four different sub-forms of adaptive automation. They are based on different underlying principles to initiate a change in automation level \cite{inagaki_adaptive_2008,parasuraman_mitigating_2008}. A change may be initiated based on the following principles. (a) Operator performance is continuously assessed to decide whether the operator needs more or less support by automation. (b) Critical events (e.g., emergencies) are defined during which the operator is offered more support by automation. (c) The psychophysiological operator state (e.g. heart rate variability) is continuously monitored to determine when a change in automation is needed. (d) Operator performance models (e.g., multiple resource theory; \citeNP{wickens_signal_2000}) may be used to predict operator resource load. Considering the complexity of these questions, a lab-based simulation such as CAMS may be very helpful for carrying out empirical research to address these issues.  

Recent developments in artificial intelligence (AI) can be accommodated by the CAMS environment. CAMS allows the modelling of human-AI interaction to examine pertinent research questions (e.g., effects of IA transparency on operator trust, acceptance of recommendations by AI compared to humans). This does not necessarily require CAMS to have AI elements. It is sufficient that the human operator has the impression that there is some genuine AI operating at the other end, even though in reality it is only a simulation of AI rather than generative AI. This follows the same principle as the Wizard-of-Oz technique \cite{riek_wizard_2012}, though in a much more advanced (and invisible) form. However, future versions of CAMS may also make use of such AI elements (see also section \ref{sec:future_research}).  

CAMS represents a simulation environment that is sufficiently complex to be able to examine human behaviour in the management of complex systems. At the same time, the simulation environment is not too complicated for non-experienced participants to be able to understand its basic principles and functioning within a reasonable period (i.e. several hours of training). These reasonably short training requirements make it a suitable tool for experimental research, which contrasts with the high-fidelity simulators that require considerably longer training times.  

\section{Prior work using previous versions of TeamCAMS}

\begin{table}[htbp]
    \centering
    \caption{Main differences between previous CAMS versions}
    \label{tab:cams_versions} 
    \footnotesize
    \renewcommand{\arraystretch}{1.15}
    \begin{tabularx}{\linewidth}
        {>{\raggedright\arraybackslash}p{3.8cm}XXXXX}
        \hline & \textbf{CAMS} & \textbf{AutoCAMS} & \textbf{AutoCAMS 2.0} & \textbf{AutoCAMS 2.0+} & \textbf{TeamCAMS} \\
        \hline
        Level of automation for fault diagnosis\textsuperscript{1} & none & 1(3), 4, 6 & 1 to 4 & 1 to 6 & 1 to 6 \\
        Automation form & static & static & static & static, adaptive or adaptable & static, adaptive or adaptable \\ System faults available & 20 & 10 & 9 & 31 & 31 \\
        Source code & Visual Basic & Visual Basic & Java & Java & C++, Javascript \\
        Log file sampling rate & 10 s & NA & 1 s & 1 s & 1 s \\
        Collaboration and communication & no & no & no & no & yes \\
        Reference & \cite{hockey_effects_1998} & \cite{lorenz_automated_2002} & \cite{manzey_autocams_2008} & \cite{chavaillaz_operator_2017} & See present article \\
        \hline
    \end{tabularx}
    \begin{minipage}{\linewidth} \vspace{0.5cm}\footnotesize \textit{Note}: The level of automation for fault diagnosis differs considerably between CAMS versions, whereas the system for system stabilisation is automated in all versions and differs little between them. NA = not applicable. \textsuperscript{1} According to \cite{sheridan_human_1987}. \end{minipage}
\end{table}

\subsection{History of CAMS development}
The development of CAMS can be divided into five major phases: (1) creating the basic version of an automated system; (2) modelling three levels of automation; (3) extending the number of automation levels as well as changing the system architecture; (4) extending automation levels being modelled to six and simulating adaptive and adaptable automation as further design options; and (5) extension to a distributed system permitting the modelling of teamwork. Table \ref{tab:cams_versions} provides an overview of the main differences between the different CAMS versions. Please note that a partially simplified version of AutoCAMS was also recently developed for optimising physiological research \cite{xu_overcoming_2023}. However, we did not include it in table \ref{tab:cams_versions} since it represents a simplification rather than an advancement in developing additional functionalities, though we acknowledge that this simplified version has it merits with regard to ease of experimental usage. 

\paragraph{Stage 1 - CAMS.} CAMS was developed in the early 1990ies at the University of Hull (United Kingdom) for the purpose of creating a tool, which allowed examining the issue of maintaining performance of space crews during extended spaceflight \cite{hockey_effects_1998}. Due to the focus on spaceflight-related issues, the operational context of CAMS was embedded in a space capsule. However, it is considered a generic tool (i.e. applicable to a range of work domains). This is because CAMS models a complex process control system (i.e. in the form of spacecraft’s life support system), which shares its main system characteristics (e.g., opaqueness, closely coupled system parameters, long time lags) with systems from many other work domains (e.g. processes control in the chemical industry, power plants). A description of the original CAMS version may be found in \cite{sauer_conceptual_2000}.  

\paragraph{Stage 2 - AutoCAMS.} An important enhancement of the simulation environment was made by \cite{lorenz_automated_2002} at the Catholic University of Washington (United States), whose main contribution was the modelling of three automation levels (i.e. levels 3, 4 and 6), outlined in the model by \cite{sheridan_human_1987}. This allowed opening up a new research avenue, which permitted to examine the consequences of using certain automation levels for operator performance and subjective operator state (e.g., fatigue, mental effort expenditure). A description of this version of CAMS may be found in \cite{lorenz_automated_2002}.  

\paragraph{Stage 3 - AutoCAMS2.0.} The next step involved increasing the number of automation levels to four, including levels 1-4, as described in \cite{sheridan_human_1987}. This allows for an assessment of the impact of the different automation levels on performance and the subjective state of the operator. A major improvement of this version was of a technical nature, with the simulation environment being converted from a code written in MS Visual Basic to Java, which made the system independent from the specific operating system installed on the computer \cite{manzey_autocams_2008}. Furthermore, a modularised program code facilitated the future modification of the software. A very detailed description of this version of CAMS, developed at the Technical University of Berlin (Germany), may be found in \cite{manzey_autocams_2008}.  

\paragraph{Stage 4 - AutoCAMS2.0+.} The next major step in CAMS development was carried out at the University of Fribourg (Switzerland), enabling the simulation environment to make comparisons of static automation with more flexible forms such as adaptive and adaptable automation. These flexible forms allow automation levels to be changed at any time based on the operator’s decision (i.e. adaptable automation) or the machine’s decision (i.e. adaptive automation). With regard to adaptive automation, the machine’s decision can be based on operator performance, the operator’s psychophysiological state or predicting operator workload based on task difficulty. This raises also questions with regard to the number and type of automation levels that should be offered to operators \cite{sauer_how_2018}. Some features, like the onset of customized messages or rating scales during a trial, were also added. A description of this version of CAMS may be found in \cite{chavaillaz_operator_2017}. 

\paragraph{Stage 5 - TeamCAMS.} The fifth and latest step in CAMS development, carried out as a collaboration between the University of Fribourg (Switzerland) and the University of Applied Sciences and Arts of Southern Switzerland, refers to the current version of TeamCAMS, of which the technical details will be described below (see section \ref{sec:technical_description}). Its major advancement is a new system architecture for distributed teamwork, which can be run over the Internet. For example, this allows the use of tablets by ‘on-site operators’, which can be used for the examination of team-based decision processes in distributed teams. A more detailed outline of the different new features is provided in section \ref{sec:technical_description}. 

This overview of the technical development of CAMS over more than 30 years shows that it has constantly evolved by incorporating new research questions. Its development was driven by a number of universities from several countries. This continuous development has also allowed the (aesthetic) design of CAMS to be adapted, though it is still evident that the development had a strong engineering focus rather than a design one, due to research being its primary purpose of use. 

\subsection{Overview of research using previous versions of CAMS }
Over the last 25 years or so, a considerable number of studies have been carried out, which made use of one of the previous versions of TeamCAMS. A literature review revealed 42 articles that were published in peer-reviewed journals. In addition, there were conference papers presenting work with CAMS, which were however not included in the present article. An overview of the studies is presented in table 2. 

The overview table shows that the work using CAMS covers a wide range of research areas, with automation design, stress and training being particularly prominent topics. A major focus of research on automation design centred around the question of implementing adaptable and adaptive automation to provide better support for the operator than traditional static automation. In particular, the issue of psychophysiology-based adaptive automation gained considerable interest. Further topics were the implications of automation reliability and taskload for the work of operators. The topic stress and automation covered a range of typical work stressors, including environmental stressors (notably noise as well as isolation and confinement), organisational stressors (notably nightwork) but also social stressors (e.g. social exclusion, negative feedback). Similar to automation design as a major intervention in the area of human factors and ergonomics, various options in training design were examined to improve operator state and performance. This work covers a wide range of training methods (e.g. procedure-based, knowledge-based, heuristics, drill and practice, situation awareness training, and error training). A number of studies across the different topics also examined teamwork with CAMS or personnel selection issues related to operator characteristics. This very brief synopsis of the work demonstrates the highly versatility of this simulation environment, which could also be indicative of its potential for future work, as described in section \ref{sec:future_research}.  

\footnotesize
\begin{longtblr}[
  caption = {Overview of peer-reviewed articles using a version of CAMS as the main task (categorised by primary research areas)},
  label = {tab:cams_literature},
]{
  colspec = {p{8.0cm,t}p{4.5cm,t}p{3cm,t}},
  rowhead = 1,
  stretch = 1.0,
  rowsep = 2pt,
  hline{1,2,Z} = {0.08em},
  %row{odd} = {bg=gray!15}, 
  %row{even} = {bg=white},
  row{2,4,6,8,10,12} = {bg=gray!15}, 
  row{15,17,19} = {bg=gray!15},
  row{22,24,26,28,30,32} = {bg=gray!15},
  row{35,37,39,41,43,45,47} = {bg=gray!15},
}

\textbf{Main variables examined} & \textbf{Main topic of study} & \textbf{Reference} \\

\SetRow{bg=gray!50} \SetCell[c=3]{c}\textbf{\textsc{Area: Automation Design}} & & \\ 

Level of automation (low, medium, high) / fault presence &
Degree of automation support
&
\cite{lorenz_automated_2002}
\\

Extended non-practice of skills; level of system reliability (60\%, 80\%, 100\%)
&
Automation reliability and skill retention
&
\cite{chavaillaz_effects_2016}
\\

Level of system reliability (60\%, 80\%, 100\%) and fault scenario (10 fault states)
&
Automation reliability
&
\cite{chavaillaz_system_2016}
\\

Task load (failure of 1, 3 or 5 subsystems) and task motivation (normal vs. high)
&
Task load and motivation
&
\cite{earle_separating_2015}
\\

Level of automation support (4 levels) and exposure to automation failures
&
Degree of automation support
&
\cite{manzey_human_2012}
\\

Type of automation (manual, decision selection, action implementation) and fault state (absence vs. presence)
&
Degree of automation support
&
\cite{rottger_impact_2009}
\\

Change of level of automation in adaptable automation (levels 1--6)
&
Degree of automation support
&
\cite{sauer_how_2018}
\\

System reliability (60\% vs. 100\%), lay-off period and time-on-task (4 levels)
&
Automation reliability and skill retention
&
\cite{sauer_effectiveness_2017}
\\

Triggering criterion (level of deviation from standard performance), type of task (primary vs. secondary tasks), and baseline of performance data (e.g. moving average)
&
Performance-based adaptive automation
&
\cite{sauer_effectiveness_2017}
\\

Information processing stage (information acquisition and analysis, decision and action), level of autonomy (low vs high), agent operational condition (routine and failure)
&
Automation design
&
\cite{xu_overcoming_2023} \\ \pagebreak
\SetRow{bg=gray!50} \SetCell[c=3]{c}\textbf{\textsc{Area: Psychophysiology}} & & \\ 

Task load (number of failed systems requiring manual control: 1-5), direction of task load change (increasing vs. decreasing)
&
Task load level and direction of task-load change
&
\cite{nickel_online_2006}
\\

Psychophysiological markers of strain (EEG power ratio and heart rate variability) and task load (1-5 systems to control manually)
&
Psychophysiological markers of strain and task-load level
&
\cite{hockey_sensitivity_2009}
\\

Task load (1-4 failed systems requiring manual control)
&
Task load level, 
Operator functional state modelling
&
\cite{cao_recognition_2021}
\\

Use of psychophysiological markers and task performance for predicting functional state of operator
&
Predicting functional state of operator
&
\cite{zhang_predictive_2013}
\\

Use of psychophysiological markers (e.g. heart rate variability) and task performance for predicting functional state of operator
&
Predicting functional state of operator
&
\cite{wang_assessment_2012}
\\

Use of psychophysiological indices (e.g. heart rate, heart rate variability) and task load index for predicting functional state of operator
&
Predicting functional state of operator
&
\cite{wang_adaptive_2015}
\\ \pagebreak
\SetRow{bg=gray!50} \SetCell[c=3]{c}\textbf{\textsc{Area: Stress \& Automation}} & & \\

Source of social stress (automation, human, no stress)
&
Social stress induced by automation
&
\cite{thuillard_human_2024}
\\

Social stress using Trier Social Stress Test (stress vs. no-stress)
&
Social stress
& \cite{peifer_relation_2014}
\\

Social stress using Trier Social Stress test (stress vs. no-stress) and noise (noise vs. quiet)
&
Social and environmental stress
&
\cite{peifer_effects_2020}
\\

Automation control mode (adaptable automation, performance-based adaptative automation, event-based adaptive automation), environmental stress (noise vs. quiet), fault state familiarity (practiced, novel)
&
Environmental stress (noise) and automation control modes
&
\cite{sauer_comparison_2012}
\\

Mode of adaptable automation (free choice, prompted choice, forced choice), environmental stress (quiet vs. noise), fault state familiarity (practiced, novel)
&
Environmental stress (noise) and adaptable automation
&
\cite{sauer_explicit_2011}
\\

Automation mode (low medium and high static automation, and adaptable automation), environmental stress (quiet vs. noise),
&
Environmental stress (noise) and automation mode
&
\cite{sauer_designing_2013}
\\

Sleep deprivation (yes/no) and interface dialogue control (human-centred vs. machine-centred)
&
Sleep deprivation and level of operator control
&
\cite{hockey_effects_1998}
\\

Time of day (night vs. day), automation support (lower vs higher)
&
Night work and Degree of automation support
&
\cite{reichenbach_human_2011}
\\

Time of day (night vs. day), time-on-task, and fault type (none, practised, novel faults, highly complex fault)
&
Nightwork and fault difficulty
&
\cite{sauer_designing_2003}
\\

Duration of isolation and confinement (135-day spaceflight simulation)
&
Social isolation and physical confinement
&
\cite{sauer_maintenance_1999}
\\

Duration of isolation and confinement (8-month Antarctic wintering-over expedition)
&
Social isolation and physical confinement
&
\cite{sauer_performance_1999}
\\

Duration of isolation and confinement (7-day spaceflight simulation)
&
Social isolation and physical confinement
&
\cite{sauer_multiple-task_1999}
\\ \pagebreak
\SetRow{bg=gray!50} \SetCell[c=3]{c}\textbf{\textsc{Area: Training \& Automation}} & & \\ 

Training method (emphasis shift training, combined with situation awareness training, drill and practice), skill retention period (short, medium, and long-term), fault type (practised vs novel)
&
Training type and skill retention
&
\cite{burkolter_comparative_2010}
\\

Training (procedure-based vs. heuristics)
&
Training type
&
\cite{burkolter_predictive_2009}
\\

Type of training (emphasis shift training, situation awareness training, or drill and practice)
&
Training type
&
\cite{burkolter_assessment_2010}
\\

Training method (system-based vs. procedure-based), fault type (none, practised, novel faults, highly complex fault), lay-off period (short-term, long-term skill retention) 
&
Training type and skill retention
&
\cite{sauer_conceptual_2000}
\\

Training (system-based vs. procedure-based), environmental stress (quiet, noise, quiet), fault type (none, practised fault, novel fault, complex fault)
&
Training type, environmental stressor (noise) and fault difficulty
&
\cite{hockey_adaptability_2007}
\\

Training on heuristic rules (heuristic vs basic) / fault types (fault-free, practised faults, novel faults, highly complex faults)
&
Training type, Experience (old / new failures)
&
\cite{sauer_effects_2008}
\\

Training method (drill and practice vs. error training) \newline Covariate: individual differences (general mental abilities, cognitive style and conscientiousness)
&
Training type and operator characteristics
&
\cite{kluge_interaction_2011}
\\

Training method (drill and practice, error training, procedure-based and heuristic training), skill retention (medium vs. long lay-off period), fault state familiarity (high vs. low)
&
Training type, skill retention and fault state familiarity
&
\cite{kluge_designing_2010}
\\

Training method (heuristic rules vs basic), fault type (none, practised faults, novel faults, highly complex faults)
&
Training type, Experience (old / new failures)
&
\cite{sauer_effects_2008}
\\

Training experience (high-reliability training group, misdiagnosis-prone system group, miss-prone group), nature of automation failure (none, failure of diagnostic function, failure of alarm function)
&
Training type and nature of automation failure
&
\cite{sauer_experience_2016}
\\

Communication skills training (yes/no), congruence of competence and hierarchical status (leader more competent than assistant vs. assistant more competent than leader) 
&
Training for effective teamwork in incongruent teams
&
\cite{sauer_multi-level_2010}
\\

Type of training (low-level procedure-oriented vs. high-level knowledge-oriented training), cognitive diversity (high specialisation vs. low specialisation)
&
Teamwork and cognitive diversity
&
\cite{sauer_cognitive_2006}
\\

Type of training (warning of failure without prior experience of failure vs. failure warning without prior experience of failure)
&
Training for managing automation failures
&
\cite{bahner_misuse_2008}
\\

Study 1: procedural aid (standard vs. enhanced to reduce cognitive load) and fault types (familiar, novel, highly complex) \newline Study 2: procedural guidance on use of decision aid (standard vs. enhanced) 
&
Design of procedural training aid and type of training
&
\cite{kluge_combining_2013}
\\

\SetRow{bg=gray!50} \SetCell[c=3]{c}\textbf{\textsc{Area: Other Topics}} & & \\ 

Predictors: general mental ability (GMA), need for cognition (NC)
&
Operator characteristics and performance in process control tasks
&
\cite{burkolter_process_2012}
\\

\end{longtblr}

\normalsize
\section{Technical description of TeamCAMS}
\label{sec:technical_description}

\subsection{Main features and major enhancements}
Both from a technical point of view and from a researcher’s perspective, TeamCAMS represents a major advancement. From a technical point of view, TeamCAMS is based on a client-server architecture, which enables the collaboration of several participants on individual computers during an experimental session. The simulation itself is executed and managed by a \emph{manager} application running on the researcher’s computer, which serves as the authoritative source of the simulation state and experimental control. Participants interact with the simulation through a browser-based \emph{operator interface}, eliminating the need to install dedicated software and enabling access from a variety of devices, including desktop computers, laptops, and tablets. Because the operator interface runs in a standard web browser, TeamCAMS can be used both in laboratory settings and remotely, making it possible to provide training scenarios and experimental sessions that participants can complete from home or other off-site locations. Communication between the manager and the operator interfaces is mediated by an MQTT-based messaging infrastructure\footnote{MQTT is a lightweight communication protocol that allows distributed software components to exchange messages over a network. In TeamCAMS, MQTT is used solely as a communication layer between the manager and the operator interfaces; the simulation itself remains hosted on the researcher’s computer and does not require deployment to a cloud service.}. MQTT is used solely as a communication layer between the different components and does not require deployment of the simulation in the cloud, allowing the simulation itself and all associated data to remain under the direct control of the researcher. Furthermore, all communication between the manager and connected operator interfaces is encrypted, ensuring that experimental data and participant interactions are accessible only to authorized researchers. This architecture also enables clients and servers to operate across different networks while allowing multiple participants to interact simultaneously with the same simulation. Based on this client-server architecture, the experimenter can remotely control several experimental trials at the same time while receiving all logged data for further analysis. Furthermore, several additional tools for the experimenter were developed such as a visual script editor (i.e. facilitating the definition of simulation scenarios by presenting a graphical timeline of the study setup), which makes the simulation environment easy to configure and allows experimenters to set up studies with ease, even those involving complex scenarios. In addition, a log processor was developed allowing for automatically extracting and batch processing data from several log files. This considerably facilitates data extraction and analysis.

\begin{figure}[ht]
\centering

\resizebox{\textwidth}{!}{%
\begin{tikzpicture}[
font=\small,
manager/.style={
draw,
rounded corners,
fill=blue!10,
minimum width=3.8cm,
minimum height=1.2cm,
align=center
},
experiment/.style={
draw,
rounded corners,
fill=gray!10,
minimum width=2.4cm,
minimum height=0.9cm,
align=center
},
operator/.style={
draw,
rounded corners,
fill=green!10,
minimum width=2.3cm,
minimum height=0.9cm,
align=center
}
]

%%%%%%%%%%%%%%%%%%%%%%%%%%%%%%%%%%
% MANAGERS
%%%%%%%%%%%%%%%%%%%%%%%%%%%%%%%%%%

\node[manager] (m1) at (-5,3)
{
\textbf{TeamCAMS Manager A}\\
Researcher's computer
};

\node[manager] (m2) at (5,3)
{
\textbf{TeamCAMS Manager B}\\
Researcher's computer
};

%%%%%%%%%%%%%%%%%%%%%%%%%%%%%%%%%%
% EXPERIMENTS
%%%%%%%%%%%%%%%%%%%%%%%%%%%%%%%%%%

\node[experiment] (e1) at (-6.5,1)
{Experiment 1};

\node[experiment] (e2) at (-3.5,1)
{Experiment 2};

\node[experiment] (e3) at (3.5,1)
{Experiment 3};

\node[experiment] (e4) at (6.5,1)
{Experiment N};

\draw[thick] (m1) -- (e1);
\draw[thick] (m1) -- (e2);

\draw[thick] (m2) -- (e3);
\draw[thick] (m2) -- (e4);

%%%%%%%%%%%%%%%%%%%%%%%%%%%%%%%%%%
% OPERATORS
%%%%%%%%%%%%%%%%%%%%%%%%%%%%%%%%%%

\node[operator] (o11) at (-8.0,-4)
{
Operator 1\\
Browser
};

\node[operator] (o12) at (-5.0,-4)
{
Operator 2\\
Browser
};

\node[operator] (o21) at (-2.0,-4)
{
Operator\\
Browser
};

\node[operator] (o31) at (2.0,-4)
{
Operator 1\\
Browser
};

\node[operator] (o32) at (5.0,-4)
{
Operator 2\\
Browser
};

\node[operator] (o41) at (8.0,-4)
{
Operator\\
Browser
};

%%%%%%%%%%%%%%%%%%%%%%%%%%%%%%%%%%
% COMMUNICATION PATHS
%%%%%%%%%%%%%%%%%%%%%%%%%%%%%%%%%%

% Experiment 1 -> two operators
\draw[dashed,-,thick] (e1) -- (o11);
\draw[dashed,-,thick] (e1) -- (o12);

% Experiment 2 -> one operator
\draw[dashed,-,thick] (e2) -- (o21);

% Experiment 3 -> two operators
\draw[dashed,-,thick] (e3) -- (o31);
\draw[dashed,-,thick] (e3) -- (o32);

% Experiment N -> one operator
\draw[dashed,-,thick] (e4) -- (o41);

%%%%%%%%%%%%%%%%%%%%%%%%%%%%%%%%%%
% MQTT LAYER
%%%%%%%%%%%%%%%%%%%%%%%%%%%%%%%%%%

\node[
draw,
fill=orange!15,
minimum width=\textwidth,
minimum height=1.2cm,
align=center
] (mqtt) at (0,-1)
{
\textbf{MQTT Communication Layer}\\
Encrypted communication between simulations and operator interfaces
};

%%%%%%%%%%%%%%%%%%%%%%%%%%%%%%%%%%
% LABELS
%%%%%%%%%%%%%%%%%%%%%%%%%%%%%%%%%%

\node[font=\small] at (-6.5,-5.1)
{Desktop, laptop, tablet};

\node[font=\small] at (5.0,-5.1)
{Local or remote access};

\end{tikzpicture}
}

\caption{
Overview of the TeamCAMS architecture. Researchers can manage multiple experimental sessions simultaneously, while participants interact through browser-based interfaces on different devices. An encrypted MQTT layer connects operators and experiments across networks while keeping simulations and data under researcher control.
}
\label{fig:teamcams_architecture}

\end{figure}
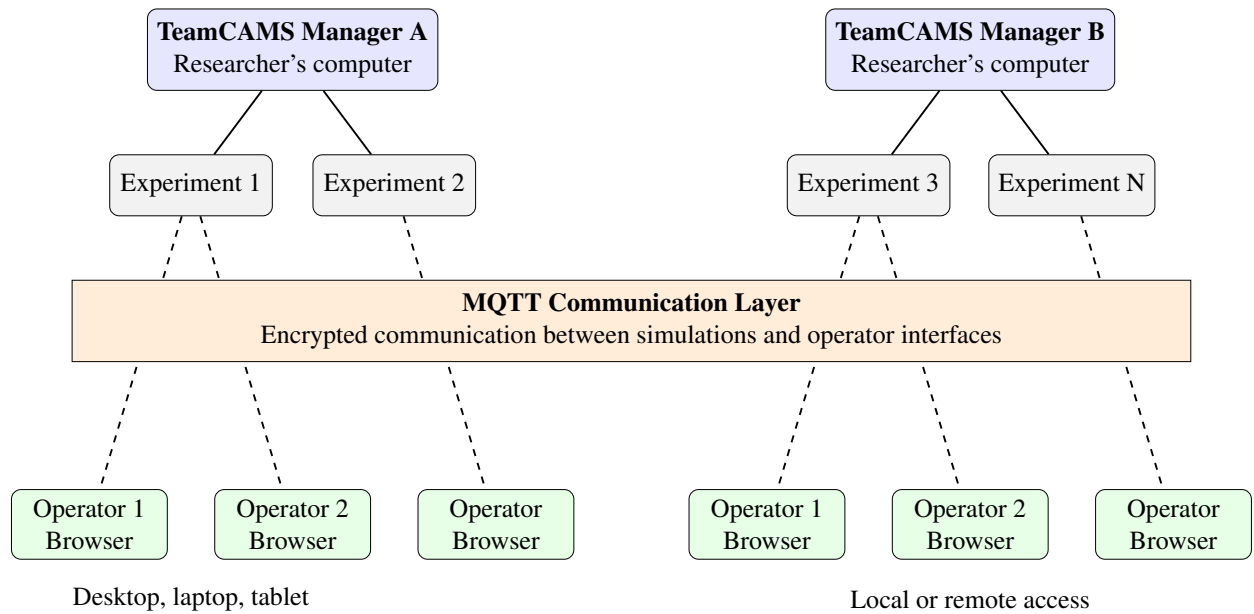

From a functional point of view, the first important change is the multi-user support allowing for a simulation of teamwork in the CAMS environment. Second, there is now a messaging system that enables participants to remotely collaborate as well as providing the opportunity to simulate communication of an artificial intelligence with participants (e.g. in the context of algorithmic leadership in which the automated system assigns tasks, gives instructions and performance feedback to participants). Third, a newly introduced questionnaire tool provides the possibility to administer questionnaires directly from the simulation. Fifth, a newly developed tutorial facilitates the training of participants. Finally, the user interface has also been improved to meet current design standards.  

Compared to previous versions of the software, the current version provides several improvements. The main changes are as follows: 
\begin{itemize}
\item \textbf{Improved interface:} It mimics physical controls (i.e. buttons, switches, displays) and provides both visual and audible feedback (i.e. alarms, notifications). 
\item \textbf{Client-server architecture:} It allows for decoupling the set-up phase from the simulation phase. Researcher can easily control simulation runs from a different device, and multiple simulations can be carried out concurrently.  
\item \textbf{Introducing multi-user support:} Multiple users can operate (or cooperate) on the same simulation. Moreover, different users can be assigned different views of the CAMS system, hence tasks can be divided among people working in a team. 
\item \textbf{Introducing multi-device support:} The simulation can be run on multiple devices at the same time, allowing for complex configurations integrating both desktop computers and mobile devices (e.g., tablets). Devices can be configured to only show a specific panel of the operator's interface, allowing for multi-screen interaction. 
\item \textbf{Introducing a messaging system:} A messaging system based on a built-in textual chat function supports direct communication between experimenters and participants as well as among participants themselves. In addition, message transmission can be parametrized to introduce controlled delivery delays, thereby allowing researchers to simulate communication latency and examine its impact on team performance, coordination processes, and collaborative decision making. 
\item \textbf{Introducing questionnaires:} Online and off-line questionnaires can be directly included in TeamCAMS, enabling experimenters to collect useful information at specific moments throughout the study. 
\item \textbf{Introducing interactive tutorials:} The initial training on the use of TeamCAMS for test participants is time-consuming and complicated due to the complexity of the task. Therefore, interactive tutorials were implemented which allow the experimenter to reduce training time for new users of the CAMS interface. Several integrated interactive tutorials are available, as well as a feature to create new ones. 
\end{itemize}

\subsection{Experimental scenario}
\begin{figure}[ht] \centering
\begin{tikzpicture}[
    scale=0.95,
    every node/.style={font=\small},
    pipe/.style={line width=4pt},
    flow/.style={->,very thick},
    equipment/.style={draw,minimum width=0.8cm,minimum height=1.2cm},
    valve/.style={draw,circle,minimum size=0.5cm},
]

% Tanks
\node[equipment,fill=blue!80] (o2tank) at (0,2) {};
\node[left=0.1cm of o2tank] {O$_2$ tank};

\node[equipment,fill=green!80] (n2tank) at (0,0) {};
\node[left=0.1cm of n2tank] {N$_2$ tank};
% Valves
\node[valve] (o2valve2) at (2.8,2) {};
\node[above=0.2cm of o2valve2] {O$_2$ valve};
\node[valve] (n2valve2) at (2.8,0) {};
\node[below=0.2cm of n2valve2] {N$_2$ valve};

% Mixer
\node[draw,circle,minimum size=0.6cm,fill=gray!20] (mixer) at (4.5,1) {$\times$};
\node[left=0.1cmof mixer] {Mixer};

% Pipes
\draw[pipe,blue] (o2tank.east)--(o2valve2)--(4.5,2)--(mixer);
\draw[pipe,green] (n2tank.east)--(n2valve2)--(4.5,0)--(mixer);

% Cabin main duct
\draw[pipe,blue] (mixer)--(4.99,1.1)--(7,1.1);
\draw[pipe,green] (mixer)--(4.99,0.9)--(7,0.9);

% Cabin
\draw[thick,gray] (7,2.2) rectangle (13,-0.2);
\node at (10,1) {\textbf{Cabin}};

% Heater
\draw[pipe,orange] (7.7,2.2)--(7.7,2);
\node[above] at (7.7,2.2) {Heater};

% Cooler
\draw[pipe,orange] (10.0,2.2)--(10.0,2);
\node[above] at (10.0,2.2) {Cooler};

% Dehumidifier
\draw[pipe,cyan] (12.3,2.2)--(12.3,2);
\node[above] at (12.3,2.2) {Dehumidifier};

% Vent
\draw[pipe,blue] (7.8,-0.2)--(7.8,0.05);
\draw[pipe,green] (8.0,-0.2)--(8.0,0.05);
\draw[pipe,red] (8.2,-0.2)--(8.2,0.05);
\node[below] at (8.0,-0.2) {Vent};

% CO2 scrubber
\draw[pipe,blue] (10.1,-0.2)--(10.1,0.05);
\draw[pipe,green] (10.3,-0.2)--(10.3,0.05);
\draw[pipe,red] (10.5,-0.2)--(10.5,0.05);
\node[below] at (10.5,-0.2) {CO$_2$ scrubber};

% Crew
\draw[pipe,blue] (12.2,-0.2)--(12.2,0.05);
\draw[pipe,red] (12.4,-0.2)--(12.4,0.05);
\draw[pipe,cyan] (12.6,-0.2)--(12.6,0.05);
\draw[pipe,orange] (12.8,-0.2)--(12.8,0.05);
\node[below] at (12.5,-0.2) {Crew};

% Flow arrows
\draw[flow,white] (2,2)--(2.4,2);
\draw[flow,white] (2,0)--(2.4,0);
\draw[flow,white] (5.5,1)--(6.5,1);

% Legend
\begin{scope}[yshift=-2.0cm,xshift=-1.0cm]

\draw[pipe,blue] (0,0)--(1,0);
\draw[flow,white] (0.2,0)--(0.8,0);
\node[right] at (1,0) {O$_2$};

\draw[pipe,green] (3,0)--(4,0);
\draw[flow,white] (3.2,0)--(3.8,0);
\node[right] at (4,0) {N$_2$};

\draw[pipe,red] (6,0)--(7,0);
%\draw[flow,white] (6.2,0)--(6.8,0);
\node[right] at (7,0) {CO$_2$};

\draw[pipe,cyan] (9,0)--(10,0);
%\draw[flow,white] (9.2,0)--(9.8,0);
\node[right] at (10,0) {Humidity};

\draw[pipe,orange] (12,0)--(13,0);
%\draw[flow,black] (12.2,0)--(12.8,0);
\node[right] at (13,0) {Heat};

\end{scope}

\end{tikzpicture}
\caption{Schematic overview of the TeamCAMS life-support simulation. Oxygen and nitrogen are mixed and supplied to the cabin, while interconnected subsystems regulate environmental conditions. Participants monitor system states, diagnose failures, and take corrective actions to maintain stable operation.}
\label{fig:cabin_devices}
\end{figure}
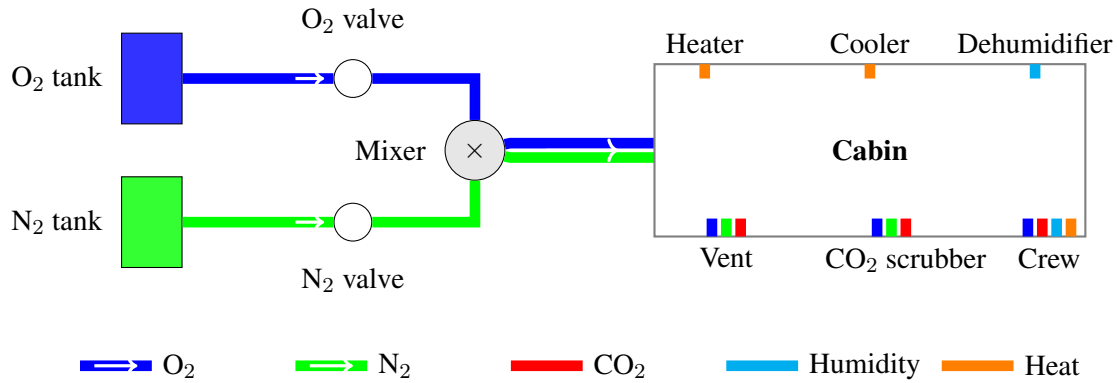

Study participants are instructed to work in a control room, monitoring a spacecraft. The simulated environment reproduces the life support system of a spacecraft. To maintain a survivable atmosphere in the cabin for the crew, the cabin is connected to an oxygen and nitrogen supply; meanwhile, a carbon dioxide scrubber ensures that the concentration of that gas remains underneath a predefined safe threshold. Furthermore, both a heater and a cooler as well as a dehumidifier maintain a comfortable condition in the cabin. A security vent is available to prevent excessive pressure in the cabin. An overview of the system, the aforementioned devices and their interaction with the cabin is shown in Figure \ref{fig:cabin_devices}: a similar schema is also displayed in the operator’s interface. 

The atmospheric conditions in the cabin environment depend on different factors, such as the concentration of oxygen, nitrogen, carbon dioxide, humidity, pressure and temperature. The simulation model mimics the one found in the original AutoCAMS 2.0 tool: 

\begin{itemize}
    \item the flow of oxygen and nitrogen into the cabin affect both the pressure and the temperature (gas flowing into the cabin reduces temperature). 
    \item the crew consumes oxygen and produces carbon dioxide, heat (positive heat flow) and humidity (positive humidity flow). 
    \item the heater increases the temperature in the cabin (positive heat flow). 
    \item the cooler decreases the temperature in the cabin (negative heat flow). 
    \item the CO$_2$ scrubber removes carbon dioxide from the cabin, as well as some oxygen and nitrogen. 
    \item the vent removes an equal amount of oxygen, nitrogen and carbon dioxide. 
    \item the dehumidifier removes humidity from the cabin (negative humidity flow). 
\end{itemize}

Since the volume of the cabin is fixed, the amount of gas and the temperature determine the atmospheric pressure. Conversely, cabin pressure determines the amount of gas that can be fed into the cabin. All simulated devices allow for manual or automatic control. In particular, it is possible to control the strength of the flow of oxygen and nitrogen fed into the cabin by acting on the corresponding valves. In a similar way it is possible to control the temperature in the cabin or the humidity level by switching on or off the heater, the cooler or the dehumidifier. Automated control relies on data gathered by sensors to trigger devices or valves on or off depending on predefined thresholds. Thresholds define two types of range: the target range, which is taken into account for automated control and should be used as reference by the operator for manual control, and the normal range, which defines the admissible value for a specific parameter (going outside this range might endanger the crew). As noted in the original AutoCAMS 2.0 documentation \cite{manzey_autocams_2008}, due to the inherent dynamics of the autonomous process control system, the system parameters may move out the target range for short times even in in the absence of a system fault. This phenomenon is due to the set point threshold that need to be passed before automated control action is taken. 

The life support system functions automatically and is automatically kept in an optimal balance. However, malfunctions or breakdowns can cause the system to lose its balance.  

\paragraph{Faults and repairs.} Components of the system might fail or operate with reduced efficiency. Faults are activated at specific times through script commands and can be diagnosed by observing the values reported for the different parameters. To keep diagnosis as simple as possible, only one type of fault can be active at a time. For each type of fault, a repair option is available by means of a repair panel, and there is a script option for clearing faults if operators cannot deal with them within the expected time period. As with AutoCAMS 2.0, faults can be grouped into five different categories: valve blockage, valve leak, valve stuck, sensor failure, and device inefficiency (see table \ref{tab:fault_examples} for some examples of system faults). Each fault influences the normal behaviour of a device and can endanger the survival of the crew. In the following, we summarize the characteristics of each simulated fault, as well as the observable symptoms, which can hint to the presence of that specific problem. For each fault the internal identifier, as used inside scripts, is also reported. 

\begin{table}[htbp]
\centering
\caption{Examples of system faults and the corresponding corrective action}
\label{tab:fault_examples}

\footnotesize
\renewcommand{\arraystretch}{1.2}
%\rowcolors{2}{gray!10}{white}
\rowcolors{2}{gray!10}{white}
\begin{tabular}{p{3.7cm}p{3.2cm}p{4.0cm}p{4.0cm}}
\toprule

\textbf{System fault} &
\textbf{Description} &
\textbf{Diagnosis} &
\textbf{Corrective action}\\
\midrule
\rowcolor{gray!10}
\textbf{Oxygen valve leak}
&
The main oxygen valve leaks some gas.
&
\vspace{-1.8em}
\begin{itemize}[leftmargin=*]
\item Concentration of oxygen in the cabin decreases.
\item Decrease rate of oxygen in the tank is higher than the oxygen flow indicated by the meter.
\item Flow of oxygen after the valve (as read on the corresponding meter) is lower than the target level corresponding to the current strength.
\end{itemize}
&
\begin{enumerate}[leftmargin=*]
\vspace{-1.8em}
\item Set oxygen feed flow to high and retain automatic mode of control.
\item Arrange for repair to be done.
\end{enumerate}
\\

\midrule
\rowcolor{white}
\textbf{Oxygen valve stuck open}
&
The main oxygen valve remains open.
&
The concentration of oxygen in the cabin increases.
&
\vspace{-1.8em}
\begin{enumerate}[leftmargin=*]
\item Turn on and off the oxygen feed flow until the level returns to the normal range.
\item Arrange for repair to be done.
\end{enumerate}
\\

\midrule
\rowcolor{gray!10}
\textbf{Scrubber ineffective}
&
The carbon dioxide scrubber efficiency is lower than normal.
&
The concentration of CO$_2$ increases.
&
\vspace{-1.8em}
\begin{enumerate}[leftmargin=*]
\item Set the scrubber strength to high and retain automatic mode of control.
\item Arrange for repair to be done.
\end{enumerate}
\\

\midrule
\rowcolor{white}
\textbf{Mixer block}
&
The mixer valve is partially obstructed and only a percentage of the oxygen and nitrogen flow passes through.
&
\vspace{-1.8em}
\begin{itemize}[leftmargin=*]
\item The cabin pressure decreases.
\item The oxygen concentration decreases.
\end{itemize}
&
\vspace{-1.8em}
\begin{enumerate}[leftmargin=*]
\item Set oxygen feed flow to high and retain automatic mode of control.
\item Set nitrogen feed flow to high and retain automatic mode of control.
\item Arrange for repair to be done.
\end{enumerate}
\\

\bottomrule
\end{tabular}

\end{table}

\paragraph{Automated fault identification and recovery agent (AFIRA).} Automation is used to support an operator in identifying and solving faults in the system. Six levels of automation are implemented, based on the taxonomy of \cite{sheridan_human_1987}. The levels of automation (LOA) provided are as follows:  
\begin{enumerate}[label=\alph*)]
\item[] \textbf{Level 1 (LOA1):} no assistance is provided. The operator is not supported by AFIRA and must diagnose and solve the problems on their own. 

\item[] \textbf{Level 2 (LOA2):} AFIRA detects the occurrence of a fault and provides a visual and an audible message; the diagnosis and resolution of the fault still need to be carried out by the operator. 

\item[] \textbf{Level 3 (LOA3):} AFIRA notifies the operator that a fault is ongoing through a visual and audible feedback, furthermore it provides a diagnosis of the problem. The operator is in charge of controlling the system and repairing the fault. 

\item[] \textbf{Level 4 (LOA4):} in addition to the support provided by the previous level of automation, a list of steps to be followed is proposed. 

\item[] \textbf{Level 5 (LOA5):} fault diagnosis and stabilisation of the system are proposed to the operator and carried out automatically after their approval (i.e. operator clicks on ‘Accept’ button). 

\item[] \textbf{Level 6 (LOA6):} AFIRA diagnoses system fault and automatically carries out fault rectification after 60 seconds of occurrence unless vetoed by the operator. 
\end{enumerate}

At LOA5 and LOA6, the automatic repair function carries out the same corrective actions as indicated in the previous section. In addition to automated fault identification and recovery, an adaptive assistance mechanism (modelling adaptive automation; \citeNP{kaber_design_2001}) has also been implemented: if enabled, the system will propose an increase or decrease of the level of automation depending on the performance of the user. Furthermore, the AFIRA system can also be disabled or configured as fallible (i.e. simulating automation failures; \citeNP{skraaning_failure_2024}), that is, it provides false support to the operator (e.g., by diagnosing a non-existent fault or by proposing the wrong diagnosis). Both adaptive automation and automation failure represent important research issues that can be examined with TeamCAMS. 

\begin{figure}[ht]
\caption{TeamCAMS operator interface showing the main information and control elements used for system monitoring, fault diagnosis, manual control, secondary tasks, automation support, and communication with other users.}
\centering
\includegraphics[width=\textwidth]{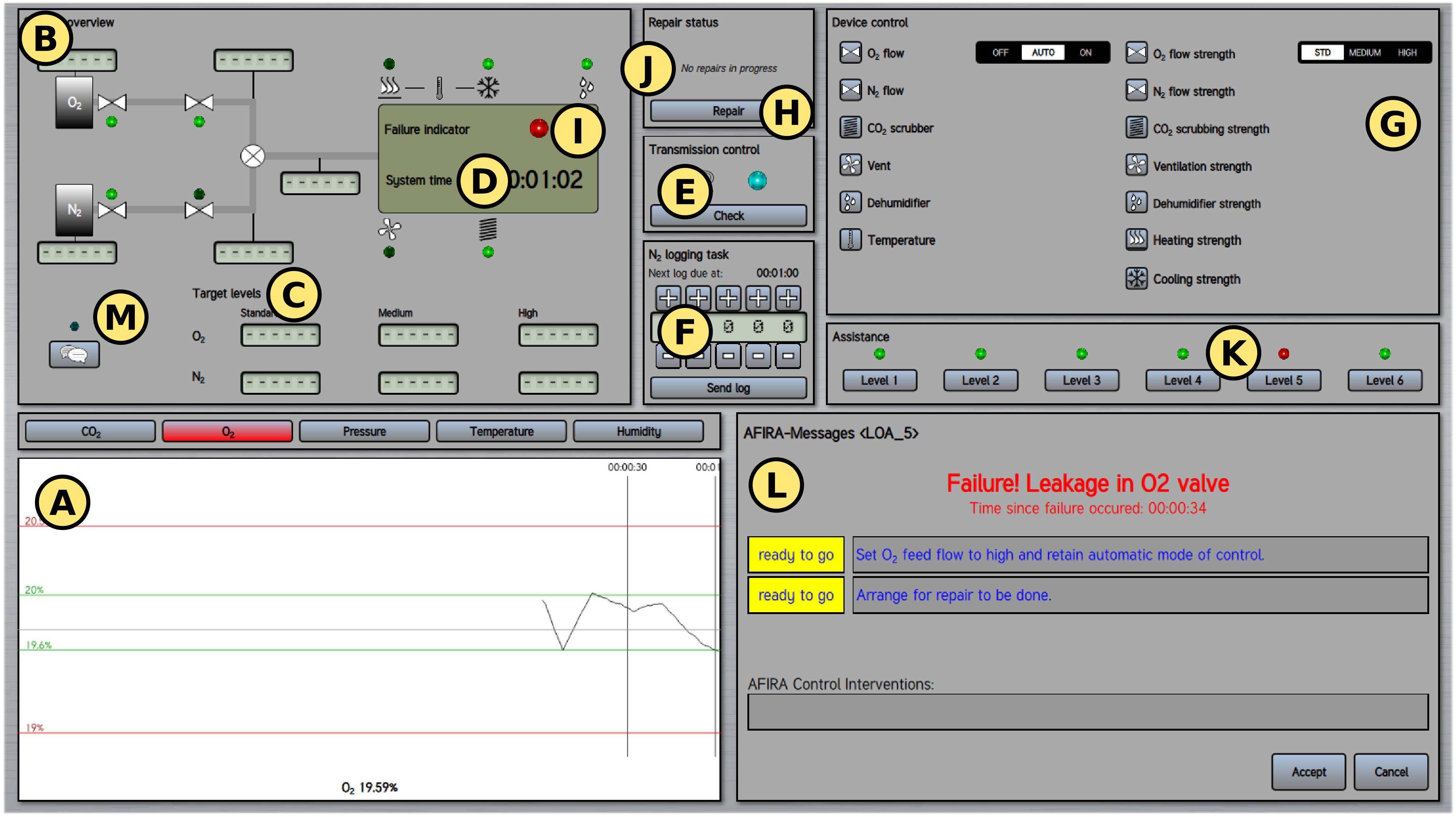}
\label{fig:operator_interface}
\end{figure}

\subsection{Operator interface}
The operator interface displays the current state of the system and allows for device control and interaction with other users. The client is an independent application, which can be executed on a remote device (e.g., desktop computer, laptop or tablet). Several clients can be used within the same experiment, either to allow several participants to be tested or to display different information on separate devices.

The operator’s interface is divided into different areas, as shown in Figure \ref{fig:operator_interface}: \textbf{(A)} history display providing the graphs of the five key parameters; \textbf{(B)} system layout with a functional schema of the cabin (including the state of specific devices such as valves and the readings of flow meters); \textbf{(C)} target values for the flow rates; \textbf{(D)} current system time; \textbf{(E)} signal of transmission control indicator and response button (for the secondary task called probe detection); \textbf{(F)} panel  for nitrogen logging task (for the secondary task called prospective memory); \textbf{(G)} control panel for adopting manual system control; \textbf{(H)} repair request button; \textbf{(I)} signal indicating system failure; \textbf{(J)} repair status indicator; \textbf{(K)} level of automatic support indicator; \textbf{(L)} automated Fault Identification and Recovery Agent (AFIRA); and \textbf{(M)} button to access the chat function. In the system overview \textbf{(B)}, green signals indicate device activity and open valves. Readings of the tank levels and flow meters are hidden by default. The participant needs to click (or tap) on the corresponding display to see the current reading, which allows for a subsequent analysis of information sampling behaviour. This action is logged, and the display disappears after 15 seconds.  

\paragraph{Display of cabin parameters }
The display, located on the bottom left of the operator’s interface (Figure \ref{fig:operator_interface}, A) shows for each cabin parameter to be monitored by the operator the current state as well as historical data of the past 240 seconds. The five different cabin parameters are carbon dioxide, oxygen, pressure, temperature and humidity. The user can temporarily display one graph by clicking on the corresponding button in the selector: the selected information remains visible for 15 seconds. Each second, a new data point is added on the right end of the graph, leading to a dynamically changing graphical representation of the status of one cabin parameter. Green guidelines in each graph indicate the target range, whereas the red lines indicate the normal range. The current value is shown at the bottom of the graph. 

\begin{figure}[ht]
\caption{Example of a device control switch. Operators can override automation by selecting one of three control modes: OFF, AUTO, or ON. Switching actions are logged for subsequent analysis of operator behaviour.}
\centering
\includegraphics[width=0.5\textwidth]{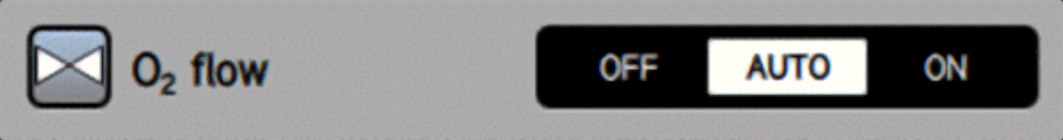}
\label{fig:switch}
\end{figure}

\paragraph{Control panel for subsystems}
The system is controlled automatically but manual control of subsystems can be adopted by the operator (see Figure \ref{fig:operator_interface}, G). Automation controls the gas flow, temperature, humidity and other cabin parameters by opening and closing valves or by operating switches according to upper and lower thresholds. The device control panel allows for overriding automation and performing manual control of each device. Manual control is especially important during fault states, because the automated control might fail or provide a false diagnosis. The device control panel is located on the top right of the interface. Control switches are normally hidden: the user can click on the desired control parameter to reveal the corresponding switch (see for instance Figure \ref{fig:switch}). The switch has three positions, OFF, AUTO, and ON. When OFF is selected, the subsystem is disabled. When ON is selected, the subsystem remains continuously active, whereas in AUTO mode, the simulation automatically controls the subsystem’s operation. The switch will remain visible for about 15 seconds. Revealing a switch is logged and can be used by the experimenter to determine the actions of the operator during an experiment. 

\paragraph{Control panel for messages}
In the control panel for messages, participants receive scripted information and instructions from the experimenter as well as messages from other participants. Furthermore, the control panel for messages contains a text entry field in order to reply to messages or to start a new conversation. It can be toggled on and off by a button (Figure \ref{fig:operator_interface}, M). All communication is logged and can be processed further by the experimenter. Furthermore, experiment scripts can trigger automatic messages to an operator. Depending on the configuration chosen by the experimenter, the dialogue is accessible either by clicking on the corresponding button in the system overview panel (Figure \ref{fig:operator_interface}, M) or by switching between the assistance panel and the control panel for messages. In the first case, an overlay dialogue will be displayed on the client screen, whereas in the second case the messaging system will be shown on the bottom right panel of the interface, replacing the assistance panel. The user can switch between the assistance and the control panel for messages by means of a slider button. When the control panel for messages is hidden, a notification light located on top of the messaging button blinks to signal for incoming (unread) messages.  If the control panel for messages is covered by the assistance panel, the messaging label on the switch button will turn blue and a notification tone will sound. 

\begin{figure}[ht]
\caption{Example of the TeamCAMS messaging system used to support team coordination and information exchange during the diagnosis of an uncertain system anomaly.}
\centering
\includegraphics[width=\textwidth]{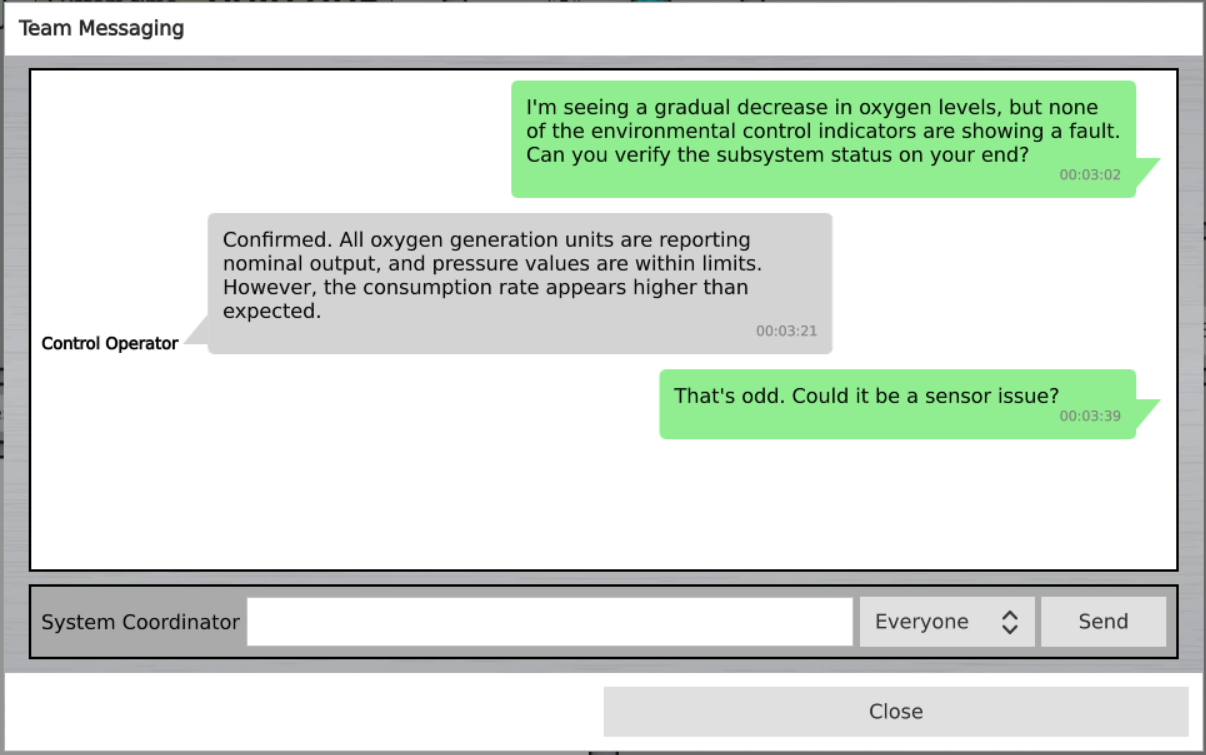}
\label{fig:messaging_window}
\end{figure}

The chat function supports both public messages as well as private conversations: private messages exchanged between users are shown in a different color. Within the messaging system users are identified by an alias instead of their username: the alias is chosen by the experimenter and allows for assigning fictional roles to each operator. 

\begin{figure}[ht]
\caption{Example of a TeamCAMS multi-device setup. Interface components are distributed across multiple devices, with controls displayed on a touch-screen device and the transmission-check task presented on a smartphone.}
\centering
\includegraphics[width=\textwidth]{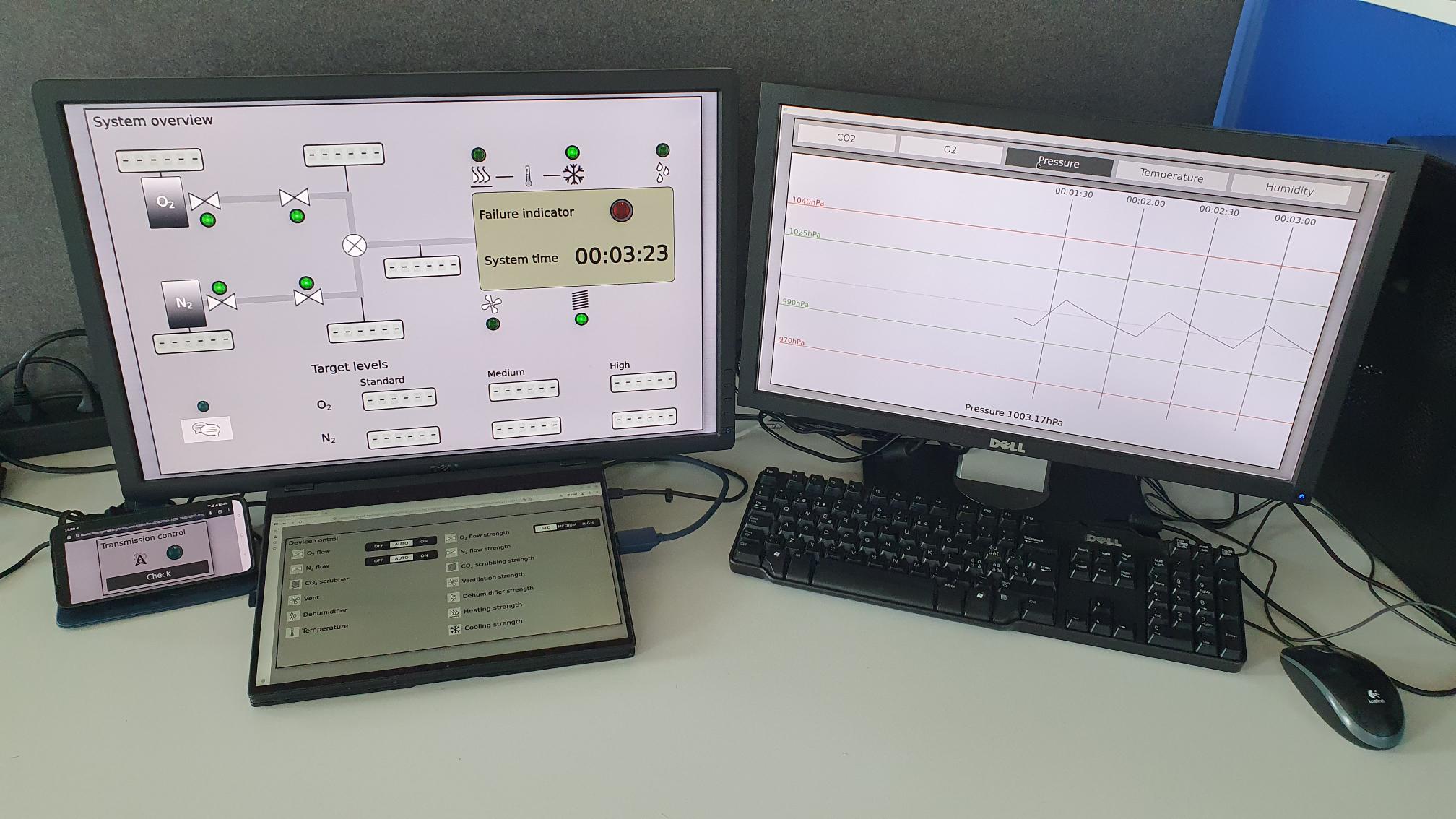}
\label{fig:operator_multidevice}
\end{figure}

\paragraph{Multi-device setup}
In addition to supporting collaboration between multiple participants within the same simulation, TeamCAMS allows the operator interface to be distributed across several devices and displays. This flexibility enables researchers to create more realistic experimental setups by separating interface components according to their functional role. For example, control panels can be presented on dedicated touch-screen devices while monitoring displays remain on desktop monitors, and secondary-task controls can be assigned to smartphones or tablets. Furthermore, different interface panels can be allocated to different team members, allowing researchers to investigate teamwork, task allocation, and collaborative decision making in complex process-control environments. Figure \ref{fig:operator_multidevice} illustrates an example of such a multi-device configuration, in which the main displays are distributed across two monitors, the control panel is operated through a touch-screen device, and the transmission-check task is presented on a smartphone. Moreover, the modular architecture of TeamCAMS allows the integration of dedicated physical interfaces and custom hardware components when required, enabling the creation of higher-fidelity experimental environments that more closely resemble real-world operational settings.

\subsection{Operator tasks} 
Participants are faced with four different tasks in this simulation, which are to be completed with different priorities. Two types of tasks are differentiated: primary tasks and secondary tasks. Primary tasks comprise system control and fault diagnosis, with the latter being influenced by the simulation script set by the experimenter. Secondary tasks include acknowledgement of transmission check alarms and periodic N$_2$ tank level logging. The importance of these tasks reflects the priorities from similar work environments found in the real world. The performance of an operator is evaluated with respect to these tasks, according to specific measurements, such as the ability to correctly solve system errors within a reasonable period. 

\paragraph{Primary task 1: system control.} A central task for an operator is to ensure that all components of the system operate correctly by observing the graphs panel (Figure \ref{fig:operator_interface}, A) and the system overview. This includes the monitoring of levels of oxygen, nitrogen, carbon dioxide as well as temperature and humidity of the space station.  

\emph{Oxygen control.} Oxygen is required to ensure that breathable air is available in the cabin to be consumed by the crew. The normal concentration range for this gas is between 19\% and 20.5\%. Oxygen flows from the corresponding tank through a control valve, then into a mixer (which combines the oxygen and nitrogen flows) and subsequently into the cabin. The main control valve is automatically opened or closed depending on the amount of oxygen in the cabin. More specifically, the valve is opened when the oxygen concentration drops below the target concentration of 19.6\% and closes when the concentration reaches 20\%. In the event of a failure, oxygen might move outside the target range. 

\emph{Nitrogen control.} The nitrogen flow is used to control the atmospheric pressure in the cabin. Therefore, the normal range is expressed in terms of millibar of pressure: the lower limit is 970 mbar, the higher limit 1040 mbar. For controlling purposes, when the pressure drops below 990 mbar the nitrogen valve is opened, whereas when the pressure rises over 1025 mbar the valve is closed. 

\emph{Carbon dioxide control.} The crew breathing in the cabin consumes oxygen and produces carbon dioxide, which is toxic and must be removed by means of a scrubbing device. The device is controlled according to the following thresholds: when the concentration of gas rises above 0.6\% the scrubber is turned on, whereas when the concentration drops below 0.3\% the scrubber is turned off. The safety range for carbon dioxide is 0.1\% to 0.8\%. 

\emph{Temperature control.} Temperature is controlled by means of a heater and a cooler device. Whenever the temperature falls below 19.5 °C the heater is turned on and the cooler is turned off, whereas if the temperature raises over 22 °C the cooler is activated, and the heater is shut down. When turned on, the cooling device, respectively the heating device, take some time to reach their maximum efficiently level. Conversely, then turned off they will still heat or cool down the cabin for a couple of seconds at a decreasing efficiency for some time, simulating the behaviour expected from a real device. The normal temperature range considered safe and comfortable for the crew is between 18.5 °C and 23 °C. 

\emph{Humidity control.} The dehumidifier is used to keep humidity within the target range, 38\% to 42\%. If humidity rises over the upper threshold the dehumidifier is turned on, whereas if the value drops below the lower threshold, the device is turned off. The normal values for this parameter range from 36.5\% to 44\%. 

\paragraph{Primary task 2: fault diagnosis and system repair.} If automation is enabled, in the event of a fault an alarm will notify the operator that a problem is occurring (Figure \ref{fig:operator_interface}, L). A manual diagnosis process is nonetheless required because the automation can also be faulty and might report an incorrect fault situation. A malfunction of the system will result in either an abnormal parameter value (i.e. a value crossing either the upper or the lower thresholds of the normal range) or, in gas flows, a mismatch between the observed flow rate and the target value recommended.

\begin{figure}[ht]
\caption{TeamCAMS repair panel. The panel allows operators to inspect system components, view available repair actions, and initiate repairs. Once a repair is started, its progress is displayed through the repair-status indicator, and no additional repairs can be initiated until the current repair has been completed.}
\centering
\includegraphics[width=\textwidth]{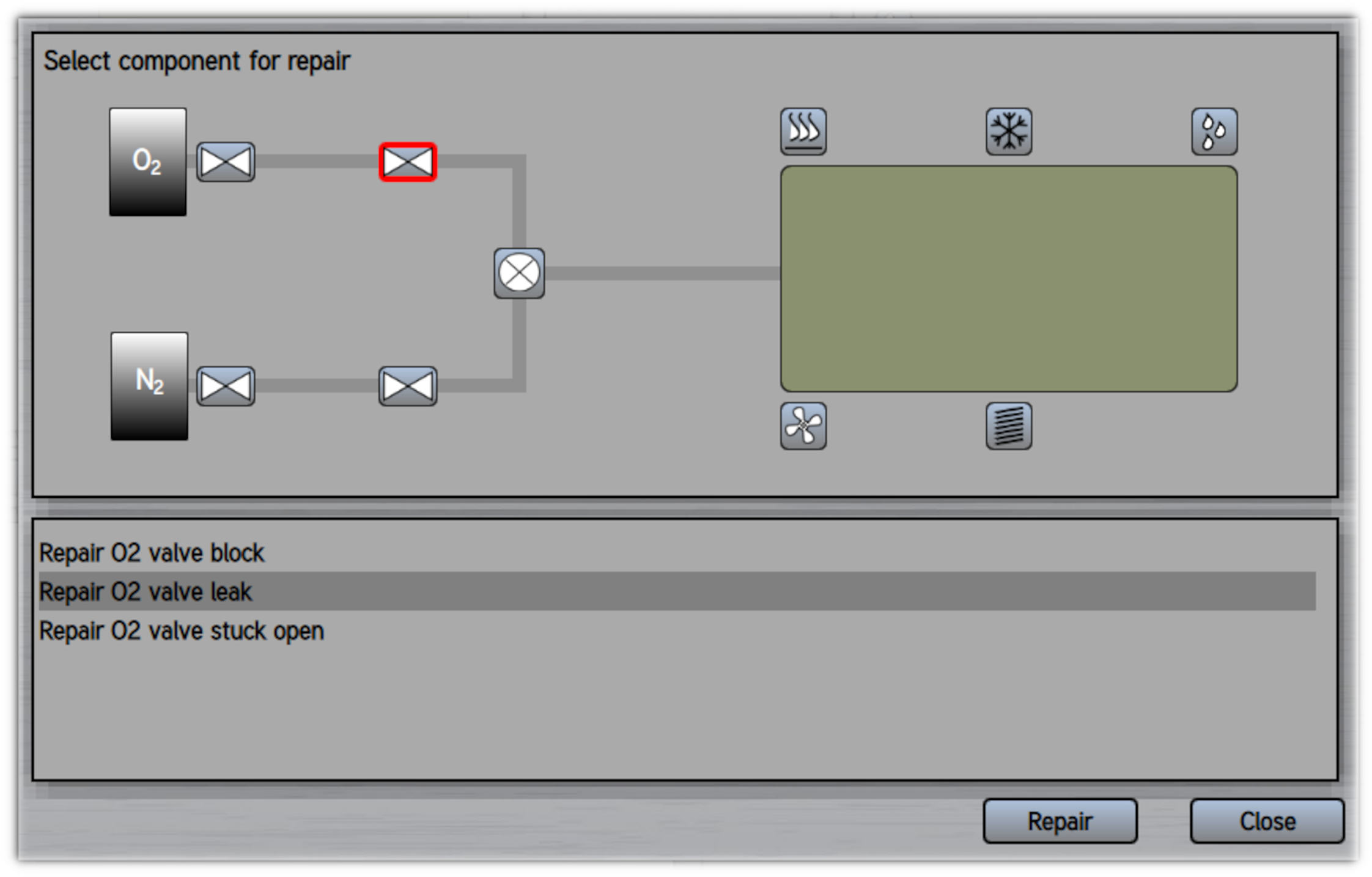}
\label{fig:repair_panel}
\end{figure}

Repairs can be performed through the repair panel, by clicking on the corresponding button (Figure \ref{fig:operator_interface}, H). A system repair panel will be shown on the operator’s interface (Figure \ref{fig:repair_panel}): the operator can click on a device to display a list of available repairs and execute the selected repair by clicking on the “Repair” button. Repairs require 30 seconds to be completed: during that time, the status of the repair is displayed on the corresponding panel (Figure \ref{fig:operator_interface}, J). Only one repair at a time is possible: if a repair operation is already in progress the repair panel is disabled. 

\paragraph{Secondary task 1: reaction time.} One of the secondary tasks is an alarm acknowledgment task, termed “connection check”, in which about every 30 seconds, the transmission check lamp (Figure \ref{fig:operator_interface}, E) will turn on, and an audible tone will be played. The operator must acknowledge this alarm by clicking on the “Check” button before the light turns off (after 10 seconds). During an experiment, the reaction time to this alarm is logged. 

\paragraph{Secondary task 2: prospective memory.} Nitrogen logging is the other secondary task to be performed by the operator. At regular intervals (e.g. about every 60 seconds) the operator needs to report the amount of nitrogen remaining in the tank. The value is to be recorded as close as possible to the due time. The submitted value as well as the submission time are logged. 

\begin{figure}[ht]
\caption{TeamCAMS Manager interface. The application provides researchers with tools to configure experimental scenarios, manage operator access, monitor ongoing sessions, and collect logged data. The Manager also serves as the central server that executes and coordinates simulation runs and communication with connected operator interfaces.}
\centering
\includegraphics[width=\textwidth]{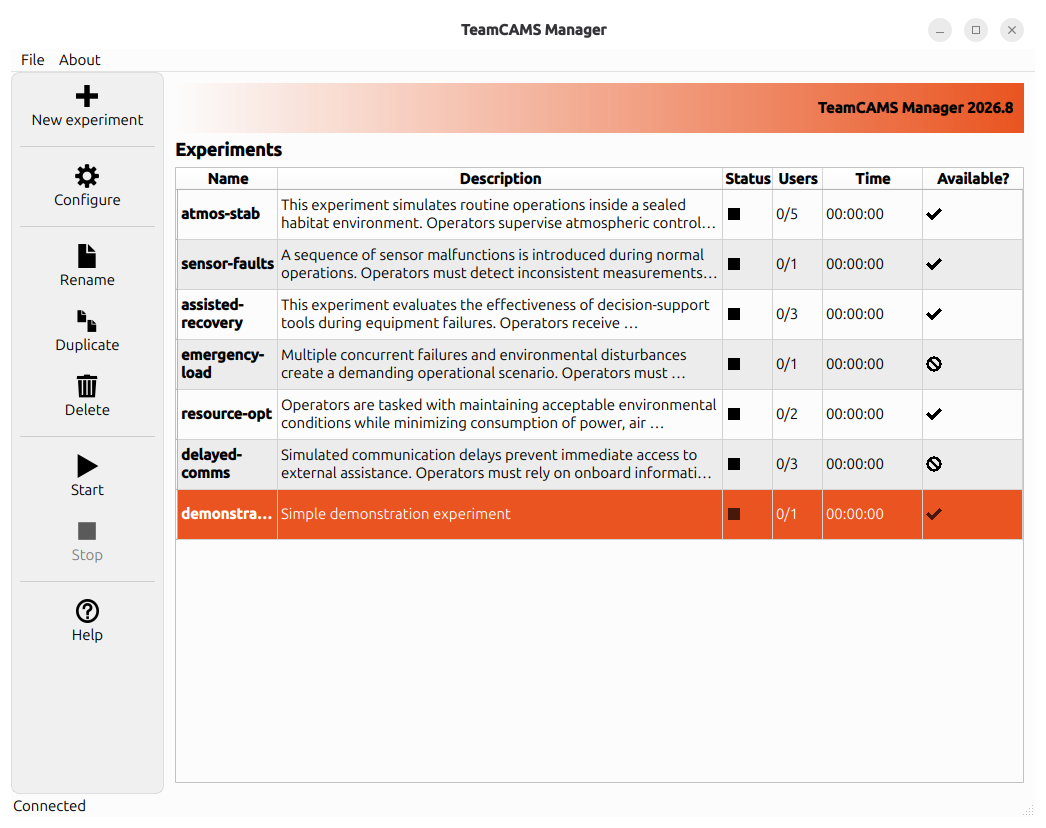}
\label{fig:manager_ui}
\end{figure}

\subsection{Setting up and running an experimental session}
\begin{figure}[ht]
\caption{TeamCAMS visual script editor used to define simulation scenarios. The graphical timeline provides an overview of the sequence and timing of scheduled events, enabling researchers to efficiently configure and manage complex experimental designs.}
\centering
\includegraphics[width=\textwidth]{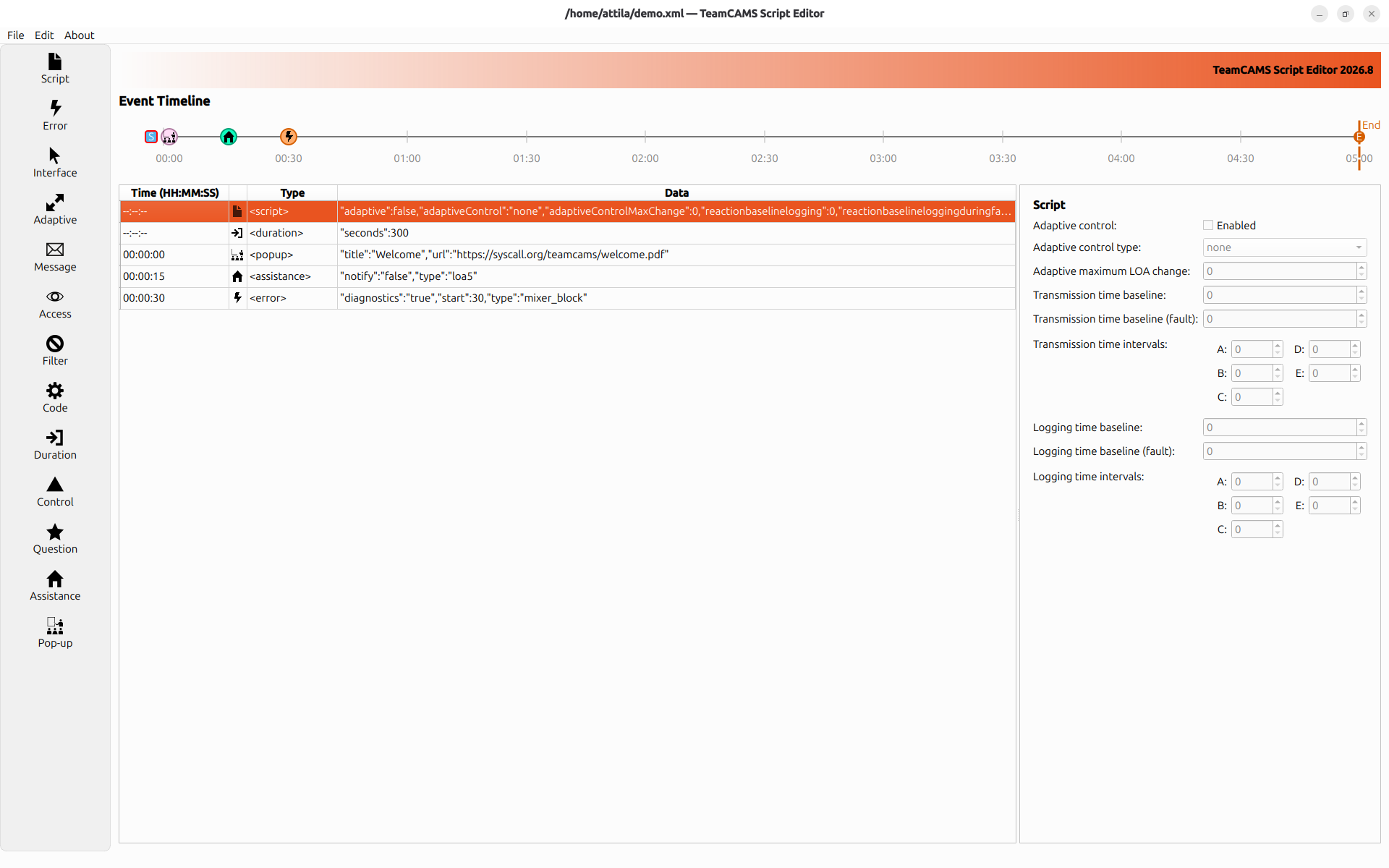}
\label{fig:script_editor}
\end{figure}

The experimenter interface contains two tools: Script Editor and Experiment Manager  (see figures \ref{fig:script_editor} and  \ref{fig:manager_ui}, respectively). The former allows for creating and editing simulation scripts, whereas the latter allows for setting up and controlling simulation runs. The Experiment Manager acts as a server for multiple operator interface clients: the experimenter can start, pause or stop experiments running either on local or remote devices at the same time. 

The events happening during a session (or experiment) can be defined using a graphical script editor (Figure \ref{fig:script_editor}). The generated XML file follows an extended syntax based on the one used by AutoCAMS 2.0. The use of a graphical tool ensures that the syntax of the generated script is correct and prevents compatibility errors.  

The experimenter can insert new events in the script by clicking on the corresponding icon on the toolbar on the left. The central part of the window shows the current timeline, whereas the right pane shows the options available for the currently selected event. Most elements have a field ‘starting time’ that defines when the event happens during the experiment. 

The script editor (Figure \ref{fig:script_editor}) allows configuring adaptive automation, which is disabled by default. Adaptive automation allows increasing or decreasing automation as a function of the operator’s  performance on the reaction time task (secondary task). Adaptive automation can be implemented in two ways. In the first approach, changes in the level of automation may be triggered if a previously specified deviation from a defined performance standard (or baseline) is detected. The experimenter can define two baselines: one for fault-free periods, and one for periods in which a fault is present. In the second approach, automation levels are linked to predefined reaction times intervals. The target level of automation is changed when the reaction time lies within certain ranges (e.g. LOA1 is selected for a score between A and B, LOA2 is selected for a score between B and C, etc.).  

\section{Future research areas and enhancements for TeamCAMS} 
\label{sec:future_research}
We have identified several interesting areas for future research, in which CAMS could be a useful research tool. This includes the following issues: designing advanced automation, remote operation, hybrid teams, and algorithmic leadership. Some of the suggested research areas may require further enhancement of the simulation environment CAMS, of which some examples are presented.  

In previous CAMS research, the design of automation was already one of the most prominent topics (e.g. \citeNP{lorenz_automated_2002, manzey_human_2012}). In future research, this classic topic in CAMS research is expected to continue to play an even more important role given the continuing technological advancements. First, further levels of automation from the classic framework of \cite{sheridan_human_1987} could be modelled in the CAMS environment. The current version only models the levels from 1-6 whereas levels 7-10 are not included in the simulation. Modelling these higher levels (in which the role of the human decreases significantly in importance) would reflect the future development of automation (see also section \ref{sec:future_research} for some suggestions). Second, another pertinent issue in automation design refers to the flexible assignment of responsibilities and the question of who should be given decision-making authority (e.g., adaptive and adaptable automation; \citeNP{inagaki_adaptive_2008}). Adaptive automation was already modelled in different forms such as physiology-based adaptive automation \cite{hockey_sensitivity_2009} or performance-based adaptive automation \cite{chavaillaz_system_2016}, though despite these efforts the research questions have not received sufficient attention given the great potential of these automation options.

Remote mobile operation simulates the control of a work system from a distance, which is facilitated by the increased availability of powerful mobile tools. There are ample examples in the field of ergonomics where operators are physically separated from the system they control. This can be a rather small distance of a few yards only (e.g. operating a crane on a building site) or a very long distance, covering thousands of miles (e.g. remote battlefield surgery, flying drones, navigating satellites). In process control, operators may move to some distant part of the plant for inspection or repair (i.e. being several hundred yards away from the control room) while still controlling the system. These forms of remote control are characterised by particular problems, such as delay in system response (i.e. increased time lags), which has long been known to be a problem \cite{rantanen_time_2004,sheridan_teleportation_1993,yang_emotional_2017}. The idea of remote control may be taken further by envisaging operators of industrial technical systems working from home. While working from home represents common practice in office jobs or call centres, such a concept of work design has hardly been implemented in the control of complex technical systems (e.g. for reasons of safety and security). However, having experienced the consequences of a recent pandemic, home-based control of complex technical systems may be an option for work design to cope with such exceptional situations. The issues of performance, system integrity, and subjective operator state during remote work compared to on-site work is worthy of some empirical research, with CAMS being able to support such research efforts.  

The social aspects of automation have been addressed in previous research (e.g., \citeNP{visser_almost_2016,domeinski_human_2007}) but we believe this is insufficient given the technological developments to be expected. These developments notably concern hybrid teams (i.e. human-machine teams involving more than one human member), which are expected to play an increasing role \cite{ellwart_i_2022}. For example, two humans and one robot are involved in an advanced manufacturing process. This also concerns social stress \cite{gerhardt_how_2021} and social support \cite{holt-lunstad_social_2015}, which may be induced and provided not only by humans but also by machines (e.g., a robot is giving negative performance feedback to an operator, followed by an algorithm providing instrumental social support to the same operator by giving advice of how to do a certain task more efficiently). An example of a research question may be the following: Does it make a difference whether the automated system or a fellow operator suggests a procedure of how to deal with a system failure? Since TeamCAMS provides an interface for a team of operators controlling the same system, aided by powerful support tools (e.g. AFIRA, support centre), it enables researchers to model the essential features of hybrid teams in a lab-based environment. 

A special and highly topical subject is automation in the specific context of leadership, that is, managing one or several members of staff. This is sometimes termed algorithmic leadership (e.g. \citeNP{cremer_humanalgorithm_2022,harms_algorithmic_2019}). In the wake of the increasing popularity of click and gig work (e.g. Amazon Mechanical Turk, Uber, etc.), the topic of algorithmic leadership has increasingly attracted the attention of the research community in recent years (for an overview see \citeNP{langer_future_2021}). Most of this research however refers to the use of hypothetical scenarios and vignettes in order to assess individual’s attitudes towards the automation of management and leadership functions \cite{wesche_peoples_2024}. While this methodological approach has its advantages (e.g. simple and economical manipulation of technological conditions without having to develop a prototype), this method is also associated with some limitations. It has been criticised that vignette-based research tends to capture attitudes and prejudices, as it is often difficult for participants to put themselves into a hypothetical situation \cite{eifler_evaluating_2007,glikson_human_2020}, that vignette studies often do not allow for the objective assessment of performance and work behaviour. Therefore, other methodological approaches are needed to evaluate algorithmic leadership systems. TeamCAMS represents a possible solution since it offers the possibility to manipulate leadership behaviour (e.g. leadership style) by means of the communication function (i.e. chat facility). This would enable us to emphasise the importance of testing such systems and evaluating the (potentially negative) consequences of their deployment before they are implemented in practice.   

For most of the research domains suggested, the current version of TeamCAMS may be suitable as a research tool. However, we would also like to suggest some specific enhancements of the simulation environment, which may be useful for future research. First, TeamCAMS could be enhanced to simulate more than the first six levels from the \cite{sheridan_human_1987} taxonomy. For example, modelling level 8 (i.e. machine informs the human only if human asks) could be of considerable interest because it allows us to determine under what circumstances the human operator is interested in receiving such information (see previous chapter). Second, the integrated messaging system (currently available in written form) could be enhanced to an auditory system for within-team communication. This would make it more realistic regarding typical within-team communication while reducing the demands on the generally very active visual system of the human. Making the TeamCAMS tool available to the research community is expected to facilitate and stimulate its future advancement. 

\section{Conclusion} 
A major goal of this article was to outline the potential of TeamCAMS to model work environments that use complex technical systems. We have outlined a number of research domains in which TeamCAMS is likely to be of some benefit. It may not only be relevant for teamwork and factors that are associated with it (e.g., social stress) but also for applications outside this domain (e.g. automation research, impact of physical stressors). This wide range of applications demonstrates that lab-based research using suitable simulation environments provides an important complement to the field-based work in these domains. 

To maintain and hopefully increase the capability of this simulation environment, several modifications have been made over recent years. As an important part, the usability of TeamCAMS was increased by making modifications to the interface, relying on design principles such as consistency and intuitiveness. Furthermore, future research can use tablets and other mobile devices in addition to the classic PC-based operation. We hope that these additions are of interest to the research community. We think that it is an advantage to have a versatile simulation environment with a proven track record over many years for reliability and continuous development over many years. 

\section*{Software Availability}
TeamCAMS is distributed as open-source software under the GNU General Public License v3 (GPLv3). 
Documentation, installation instructions, source code repositories, downloadable releases, and additional resources are available at \url{https://teamcams.syscall.org}.

\bibliographystyle{apacite}
\bibliography{bibliography}

\end{document}